\documentclass[conference]{IEEEtran}

\usepackage{graphicx}
\usepackage{tabularx}

\usepackage[T1]{fontenc}
\usepackage[utf8]{inputenc}

\usepackage{csquotes}
\usepackage{nicefrac}
\usepackage[textsize=scriptsize]{todonotes}
\usepackage{booktabs}
\usepackage{xcolor}
\usepackage{tikz}
\usepackage{quantikz}
\usetikzlibrary{arrows,positioning} 
\usetikzlibrary{shapes.geometric, positioning}
\usepackage{flushend}
\usepackage{hyperref}
\usepackage[detect-all=true]{siunitx}
\usepackage{etoolbox}
\usepackage{soul}
\usepackage{amssymb}
\usepackage{booktabs} %
\usepackage{csvsimple}
\usepackage{amsthm}
\usepackage{lipsum}
\usepackage{varwidth}
\usepackage{utfsym}
\usepackage{fontawesome5}

\robustify\bfseries

\makeatletter
\let\MYcaption\@makecaption
\makeatother
\usepackage[font=footnotesize]{subcaption}
\makeatletter
\let\@makecaption\MYcaption
\makeatother

\usepackage{yquant}
\usepackage{pgfplots}

\usepackage[style=ieee, maxnames=6, minnames=1, date=year, doi=false,isbn=false,backend=biber]{biblatex}
\usepackage[detect-all=true]{siunitx}
\usepackage{etoolbox}
\robustify\bfseries

\usetikzlibrary{fit, quotes}
\usepackage{bbm}

\newtheorem{example}{Example}

\usetikzlibrary{automata,arrows}

\AtEveryBibitem{
  \clearname{editor}%
  \clearfield{series}%
  \clearfield{isbn}%
  \clearfield{issn}%
  \clearfield{volume}%
  \clearfield{number}%
  \clearfield{pages}%
  \clearfield{doi}%
}

\newcommand{\pipath}{\boldsymbol{\pi}}
\newcommand{\pipathi}{\boldsymbol{\pi}^{(i)}}
\newcommand{\pipathN}[1]{\boldsymbol{\pi}^{(#1)}}
\newcommand{\bx}{\boldsymbol{x}}

\newcommand{\revised}[1]{#1}

\DeclareMathOperator*{\argmin}{arg\,min}

\begin{document}

\title{A Framework to Formulate \\ Pathfinding Problems for Quantum Computing}

\author{
	\IEEEauthorblockN{Damian Rovara\IEEEauthorrefmark{1}\hspace*{1.5cm}Nils Quetschlich\IEEEauthorrefmark{1}\hspace*{1.5cm}Robert Wille\IEEEauthorrefmark{1}\IEEEauthorrefmark{2}}
	\IEEEauthorblockA{\IEEEauthorrefmark{1}Chair for Design Automation, Technical University of Munich, Germany}
	\IEEEauthorblockA{\IEEEauthorrefmark{2}Software Competence Center Hagenberg GmbH (SCCH), Austria}
	\IEEEauthorblockA{\href{mailto:damian.rovara@tum.de}{damian.rovara@tum.de}\hspace{1.5cm}\href{mailto:nils.quetschlich@tum.de}{nils.quetschlich@tum.de}\hspace{1.5cm} \href{mailto:robert.wille@tum.de}{robert.wille@tum.de}\\
	\url{https://www.cda.cit.tum.de/research/quantum}
	}
}

\maketitle

\begin{abstract}
  With the applications of quantum computing becoming more and more widespread, finding ways that allow end users without experience in the field to apply quantum computers to solve their individual problems is becoming a crucial task. However, current optimization algorithms require problem instances to be posed in complex formats that are challenging to formulate, even for experts. In particular, the \emph{Quadratic Unconstrained Binary Optimization (QUBO)} formalism employed by many quantum optimization algorithms, such as the \emph{Quantum Approximate Optimization Algorithm} (QAOA), involves the mathematical rewriting of constraints under strict conditions.
  To facilitate this process, we propose a framework to \emph{automatically} generate QUBO formulations for pathfinding problems. This framework allows users to translate their specific problem instances into formulations that can be passed directly to quantum algorithms for optimization without requiring any expertise in the field of quantum computing. It supports three different encoding schemes that can easily be compared without requiring manual reformulation efforts. The resulting QUBO formulations are robust and efficient, reducing the previously tedious and \mbox{error-prone} reformulation process to a task that can be completed in a matter of seconds. In addition to an open-source Python package available on \href{https://github.com/cda-tum/mqt-qubomaker}{https://github.com/cda-tum/mqt-qubomaker}, we also provide a graphical user interface accessible through the web (\href{https://cda-tum.github.io/mqt-qubomaker/}{https://cda-tum.github.io/mqt-qubomaker/}), which can be used to operate the framework without requiring the end user to write any code.
\end{abstract}


\section{Introduction}\label{sec:intro}
The task of finding a path in a directed graph, subject to a set of constraints, is a problem class that has been thoroughly investigated for its complexity and utility. These \emph{pathfinding} problems appear in a large number of application scenarios, and finding exact or approximate solutions to them is a crucial component to solving many problems in the real world. 

While efficient classical exact algorithms exist for many pathfinding problems~\cite{hart1968, dijkstra2022}, a large number of pathfinding problems, such as the \emph{Traveling Salesperson Problem (TSP)}, the \emph{Sequential Ordering Problem (SOP)}, or the \emph{Multi-Agent Pathfinding Problem}, still lack approaches that provide satisfactory results for real-world problem instances~\cite{standley2010, escudero1988, erdem2013, lenstra1975}. Thus, classical methods for approximating the optimal solution to such problems have been an essential area of research for decades. However, these classical solutions have only been subject to minor incremental improvements in recent years.

Quantum computing solutions, on the other hand, offer the possibility of revolutionary improvements, as new and rigorously revised algorithms are still regularly proposed for the solution of many optimization problems.

In particular, while many such quantum algorithms exploit different optimization approaches, a large number of them are based on the \emph{Quadratic Unconstrained Binary Optimization~(QUBO,~\cite{kochenberger2004, kochenberger2014})} framework. Such algorithms can tackle individual optimization problems by first formulating them as QUBO problems.

As research in the field of quantum computing is gradually improving with respect to both physical devices and logical algorithms, and their accessibility becomes more and more widespread through service providers such as IBM or Amazon Web Services, interest in the application of quantum algorithms to solve a large variety of different problems is spreading rapidly to an equally diverse range of users. However, while these users may have deep knowledge in their respective domains, they are not necessarily guaranteed to be as experienced in the area of QUBO or quantum computing, \revised{and, thus, the development of tools (such as, e.g., proposed in~\cite{quetschlich2023}) that aid end users by shielding them from concepts requiring deeper knowledge of quantum computing has grown in relevance.}

Unfortunately, the process of solving individual problems with existing quantum optimization tools is not always straightforward. Input formats for quantum algorithms, such as the QUBO formalism, require extensive mathematical rewriting, a challenging task even for experts already experienced with it. While tools exist that aid the user in this process, they are not able to fully shield them from all the required expertise and still pose significant challenges for many classes of problems (this is discussed in more detail later in Section~\ref{sec:motivation}).

Due to the complex nature of pathfinding problems, this problem class suffers particularly from these challenges. Often consisting of multiple independent constraints and being tied to a specific graph structure, the desired QUBO formulation easily explodes in complexity, making their application virtually unfeasible for users without extensive expertise.

To solve these problems, this work proposes a novel solution for the \emph{automatic} generation of QUBO formulations of pathfinding problems. We propose a framework that allows end users to define problem instances through the conceptual descriptions of their constraints. These constraints are then individually translated into their corresponding QUBO formulations and finally combined into a single QUBO cost function.

This approach effectively shields users from any required expertise in the field of quantum computing, as the resulting QUBO cost functions can be translated directly into quantum circuits for the solution of individual problems. Furthermore, the framework allows the construction of the result in multiple formats. This allows end users to not only use the generated QUBO formulations in quantum algorithms but also apply them to classical QUBO solvers for comparison and evaluation purposes.

Finally, the framework also supports a total of three different encoding schemes that determine how the individual constraints are translated into intermediate quadratic cost functions. While, typically, the evaluation of different encodings requires the QUBO formulation to be remade from scratch, the proposed framework allows users to switch between the different encodings with no additional effort.

\revised{The remainder of this work is structured as follows: The core concepts of pathfinding problems and QUBO formulations, as well as related work in automatic QUBO construction, is reviewed in Section~\ref{sec:background}. Section~\ref{sec:motivation} then highlights bottlenecks and challenges in the current process of QUBO~formulation, motivating the application of an automatic framework for the task and presenting its general idea. Based on that, the proposed framework is described in more detail in Section~\ref{sec:proposed-framework}. Following this, Section~\ref{sec:casestudy} demonstrates the use of the proposed framework in two case studies and further discusses its applications. Finally, Section~\ref{sec:conclusion} proposes possibilities for future work and concludes this work.}

\section{Background}\label{sec:background}

To keep this work self-contained, this section reviews the \emph{pathfinding problem} considered throughout this work. Additionally, the workflow to solve these kinds of problems with quantum computing \revised{by representing them through QUBO~formulations} is outlined and related work in the area is discussed.

\subsection{Pathfinding Problems}
Pathfinding problems constitute a problem class that contains a broad range of various graph-based optimization problems with a myriad of applications in the real world, including logistics, scheduling, routing, artificial intelligence, and many more\mbox{~\cite{standley2010, erdem2013, lenstra1975}}.

More formally, a pathfinding problem is described by a directed graph $G = (V, E)$ with a set of vertices $V$ and a set of edges $E$. Each edge $(u, v) \in E \text{ with } u, v \in V$ is assigned a weight $w(u, v) \in \mathbb{R}$. A path $\pipath$ is defined as a sequence of vertices $(v_{i_1}, v_{i_2}, ..., v_{i_n}) \in V^n$
such that $(v_{i_j}, v_{i_{j+1}}) \in E$
for all $j \in [1, n-1]$. 
Based on this, the weight of path $\pipath$ is defined as $w(\pipath) = \sum_{(v_{i_j}, v_{i_{j + 1}}) \in \pipath}w(v_{i_j}, v_{i_{j + 1}})$. A pathfinding problem is then given by (1) a set of constraints $\mathcal{C}$ \revised{on the graph $G$}, \revised{(2) an objective function to determine if the total weight should be minimized or maximized}, and (3) the task to find a path that satisfies all constraints \revised{and optimizes the objective function}.

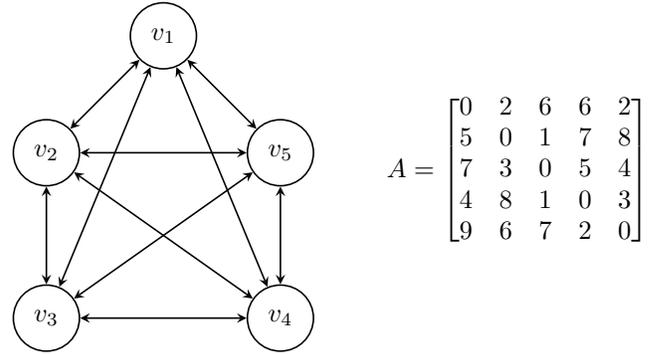
\begin{figure}
    \begin{subfigure}{0.5\columnwidth}
        \centering
        \begin{tikzpicture}[baseline=(current bounding box.center),
            > = stealth, %
            shorten > = 0pt, %
            auto,
            node distance = 2.2cm, %
            semithick, %
            ]
            \node[state] (v1) {$v_1$};
            \node[state] (v2) [below left of=v1] {$v_2$};
            \node[state] (v3) [below of=v2] {$v_3$};
            \node[state] (v5) [below right of=v1] {$v_5$};
            \node[state] (v4) [below of=v5] {$v_4$};
            
        \path[<->] (v1) edge (v2);
        \path[<->] (v1) edge (v3);
        \path[<->] (v1) edge (v4);
        \path[<->] (v1) edge (v5);
        \path[<->] (v2) edge (v3);
        \path[<->] (v2) edge (v4);
        \path[<->] (v2) edge (v5);
        \path[<->] (v3) edge (v4);
        \path[<->] (v3) edge (v5);
        \path[<->] (v4) edge (v5);
        
    \end{tikzpicture}
    \end{subfigure}
    \hspace{0.51cm}
    \begin{subfigure}{0.29\columnwidth}
        \[A = \vcenter{\hbox{$
        \begin{bmatrix}
            0 & 2 & 6 & 6 & 2 \\
            5 & 0 & 1 & 7 & 8 \\
            7 & 3 & 0 & 5 & 4 \\
            4 & 8 & 1 & 0 & 3 \\
            9 & 6 & 7 & 2 & 0 \\
        \end{bmatrix}$
        
        }}\]
        
    \end{subfigure}
    \caption{A directed graph with five vertices and seven edges. Further examples in this paper are based on this graph. \revised{The adjacency matrix $A$ represented next to the graph indicates the weight of the edge $(u, v)$ as $A_{u,v}$. A weight of $0$ indicates that no edge exists from $u$ to $v$.}}
    \label{fig:graph}
\end{figure}

\begin{example} \label{ex:1} Consider the pathfinding problem known as the \emph{Traveling Salesperson Problem (TSP)}: Given a set of cities~$V$, \revised{connected by edges $E$, the weight of $(u, v) \in E$ defined by the distance between cities $u$ and $v$}, the path shall be determined that allows the traveler to visit all cities exactly once, returns to the starting city, and has minimal weight (and, hence, constitutes the shortest path). This problem can be formulated as a pathfinding problem on a graph $G$ as shown in Figure~\ref{fig:graph}, defined by the vertices $V$, where the adjacency matrix $A_{i,j} = w(i, j)$ contains the distance (and, hence, weight) between any two cities, requiring the following four constraints: (1)~$|\pipath| = |V| + 1$ \revised{(path $\pipath$ visits a number of vertices equal to the total number of vertices in the graph, plus an additional vertex to return to the start)}, (2)~$\forall v \in V: v \in \pipath$ \revised{(all vertices of the graph are included in the path)}, (3)~$v_{i_n} = v_{i_1}$ \revised{(the path starts and ends at the same vertex)}, and (4) $w(\pipath)$ is minimal.
\end{example}

To solve such problems on a quantum computer, they must be transcribed into a formulation suitable for the available quantum algorithms.

\subsection{QUBO Formulation}

\revised{A prominent method for solving TSP or other pathfinding problems using quantum computing is to translate it into a \emph{Quadratic Unconstrained Binary Optimization} (QUBO,~\cite{kochenberger2014}) problem.} This format, used by many frequently used quantum algorithms for optimization problems, allows a large number of different problems to be encoded into a single, optimizable cost function.

QUBO problems are based on a vector of binary variables~$\bx$. A cost function $C(\bx)$ can then be defined in the form \mbox{$C(\bx) = \sum_{i} c_i x_i + \sum_{i < j} c_{i,j} x_i x_j$~\cite{lucas2014}}. The optimal solution to a QUBO problem is the vector $\bx^*$ that minimizes $C(\bx^*)$. Alternatively, the problem can be formulated as a QUBO~matrix~$Q$, such that \mbox{$C(\bx) = \bx^TQ\bx$}.

Formulating combinatorial optimization problems in the framework of QUBO problems is an active area of research~\cite{lucas2014, glover2022, whitfield2012}. To define a QUBO formulation for a specific problem, one first has to select an adequate \emph{encoding} given by an encoding function $f: \mathcal{D} \rightarrow \{0, 1\}^n$ and a decoding function $g: \{0, 1\}^n \rightarrow \mathcal{D}$, where $\mathcal{D}$ is the domain of the corresponding problem. This allows the cost function to be reformulated to $C(d) = f(d)^TQf(d)$, with $d \in \mathcal{D}$.

\begin{example}
Returning to the TSP introduced in Example~\ref{ex:1}, consider the task of encoding a given path $\pipath$ into its corresponding \emph{binary encoding} $\bx$. Consider \revised{an encoding $x_{v, i}$ for which $x_{v, i} = 1$ if and only if $v$ is the $i$-th vertex in the path, with $v \in V$ and $i \in [1, |V|]$.} This encoding defines $|V|^2$ binary variables and is an easy way to transform a directed path from and to its QUBO encoding
\footnote{Note that the literature typically uses a different encoding for the TSP~\cite{lucas2014, qian2023}. We have decided to use this encoding for the sake of generality, as it can be used efficiently with a vast number of problems.}.
\end{example}

Once an adequate encoding has been found, a cost function for the optimization of the problem can be constructed. In the framework of QUBO problems, as shown above, a cost function consists of a sum of terms that are at most quadratic in the use of binary variables. In cases where more than two variables must be multiplied, auxiliary variables can be introduced to keep the term quadratic~\cite{anthony2017}. 
Generally, problems defined over several constraints can be transformed into their QUBO cost functions by translating each constraint $\mathcal{C}_i$ into $C_i(\bx)$ individually and, then, constructing $C(\bx) = \sum_i P_i C_i(\bx)$, such that $C(\bx)$ is minimized if and only if all constraints~$\mathcal{C}_i$ are satisfied. Here, $P_i$ are penalty factors chosen so that the penalty from any violation of an individual constraint will always dominate over the potential gain by such an assignment for other constraints.

\begin{example} 
The four constraints required for the TSP listed in Example~\ref{ex:1} can be translated into a set of equivalent binary cost functions, as shown in Table~\ref{tab:cost-functions}. Here, the constraints~(1) and (2) can be encoded by a single cost function that we will call~$C_1(\bx)$. The optimization criterion (4) of finding the shortest path is represented by the cost function $C_2(\bx)$, which results in a higher penalty value the larger the total weight of the path is. Further, $x_{v, |V| + 1}$ is defined to wrap back to $x_1$. This way, it can be assumed that constraint~(3) holds implicitly. Finally, the overall cost function is constructed as \mbox{$C(\bx) = aC_1(\bx) + C_2(\bx)$}, selecting the scaling factor~${a = |V| \max_{i,j} A_{i, j}}$.
\begin{table}[]
    \caption{Cost functions that encode the constraints required for TSP.}
    \label{tab:cost-functions}
    \centering
    \renewcommand{\arraystretch}{2}
    \begin{tabular}{l l}
        \hline
    Constraint                                                                                   & Cost Function $C_i(\bx)$                                                                        \\ \hline
    \begin{tabular}[c]{@{}l@{}}
    (1) $|\pipath| = |V|$\\ (2) $\forall v \in V: v \in \pipath$\end{tabular} & $\sum\limits_{v \in V} \left( 1 - \sum\limits_{1 \leq i \leq |V|} x_{v, i}\right)^2$     \\ \hline
    (4) $w(\bx)$ is minimal                                                              & $\sum\limits_{(u, v) \in E} \sum\limits_{1 \leq i \leq |V|} A_{u,v} x_{u, i}x_{v, i +1}$  \\ \hline
    \end{tabular}
\end{table}
\end{example}

Once a QUBO problem is defined, various quantum computing approaches can be used to find its optimal solution, e.g., based on variational quantum algorithms (such as the \emph{Quantum Approximate Optimization Algorithm} (QAOA,~\cite{farhi2014}), and the \emph{Variational Quantum Eigensolver} (VQE,~\cite{peruzzo2014})), \emph{Quantum Annealing}~\cite{kadowaki1998}, and \emph{Grover Adaptive Search}~\cite{gilliam2019}.

\subsection{Related Work}
\label{sec:related-work}

Due to the complexity of creating QUBO formulations from arbitrary optimization problems, finding alternative approaches to reduce the burden on the end user is an essential area of research, and a wide range of related work exists in this field.

Multiple tools have already been devised that tackle the process of automatic QUBO formulation from different angles. These tools typically allow the user to reformulate optimization problems into a single QUBO cost function or various other output formats that can be solved using classical or quantum algorithms. Representatives for such tools are
\begin{itemize}
    \item \emph{qubovert}~\cite{qubovert}: A Python tool for QUBO formulations that also supports conversion from and to constrained or higher-order optimization problems.
    \item IBM's \emph{DOcplex}~\cite{docplex}: A library for the formulation of mathematical and constraint programming models, supporting QUBO and many other types of representations.
    \item Qiskit's~\cite{qiskit} \emph{qiskit-optimization} library: Building upon DOcplex for the formulation of problems, this tool provides solutions that are tightly coupled to many of Qiskit's existing features.
\end{itemize}

A shared property of all three of these existing tools is that they focus on constructing QUBO formulations from mathematical and logical expressions, i.e., they require the end user to already provide their problem in a numeric format. Since such formulations do not always exist naturally for arbitrary optimization problems, this still requires dedicated expertise from the end user, who is usually a domain expert in their field but not in quantum computing or QUBO. For pathfinding problems, in particular, translating their constraints and corresponding graph structure into this form leads to all the same challenges as creating the QUBO formulation, still posing a significant problem as discussed next.

\section{Motivation}\label{sec:motivation}
This section motivates the methodology proposed in this work by highlighting the main challenges of manually creating QUBO formulations and devising a general idea to mitigate them.

\subsection{Current Bottlenecks and Challenges}
\label{sec:challenges}
While solving problems given with QUBO formulations is often straight-forward with existing tools and algorithms, the same cannot be said about the construction of the QUBO formulations itself: commonly, optimization problems have to be reformulated into their QUBO representations by the end users, a time-consuming and error-prone task that has to be performed for each new problem to be considered, and which, even with the help of existing software tools that may automate individual steps of the process, suffers from multiple challenges:

\begin{enumerate}
    \item \revised{In many cases, the expertise of the end user is limited to the domain of the particular optimization problem to solve rather than the full area of QUBO formulations. Thus, translating individual problems into the QUBO formalism may be difficult or not feasible at all for a large number of users.}
    \item As a math-heavy task, the large number of binary variables required for many problems can easily cause confusion, and the interaction of variables in the cost functions can have unwanted effects. \revised{Hence, the entire process is prone to errors that are usually difficult to detect and correct at later stages.}
    \item Only quadratic terms can be considered in the setting of QUBO problems. For instance, this is shown in the cost functions listed in Table~\ref{tab:cost-functions}. Here, no term consists of a product of more than two different variables. Any terms involving three or more binary variables would have to be reformulated before being used in a QUBO problem.
    \item Only unconstrained optimization problems are part of the QUBO framework. Additional constraints must be integrated into the optimization criterion by finding fitting penalty functions as described in Section~\ref{sec:background}.
\end{enumerate}

While tools and frameworks exist that automate the translation from specific optimization problems into QUBO formulations, as reviewed in Section~\ref{sec:related-work}, they are often unable to shield end users entirely from required expertise.

\begin{example}
    \label{ex:4}
    Consider again the TSP applied to the graph depicted in Figure~\ref{fig:graph}. Creating a QUBO formulation for it with tools such as \emph{qiskit-optimization}, reviewed in Section~\ref{sec:related-work}, involves the following steps:
    \begin{enumerate}
        \item Create a quadratic programming model with a set of binary variables $e_{i, j}$ that have the value $1$ if and only if edge $(i, j)$ is in the solution graph.
        \item For each vertex i enforce the constraints $\sum_{j \in V} e_{i, j} = 1$ and $\sum_{j \in V} e_{j, i} = 1$.
        \item For each pair of vertices $(i, j)$ where $i$ is not the starting vertex, enforce the \emph{subtour elimination} constraint~${p_i - p_k + |V| * e_{i, j} \leq |V| - 1}$.
        \item Minimize the expression $\sum_{(i, j) \in E}e_{i, j}w(i, j)$.
    \end{enumerate}
    This quadratic programming model can then be translated into its QUBO formulation automatically.
\end{example}
    
Although the method reviewed above allows the user to create models with constraints, it still requires them to overcome the challenges of formulating these models mathematically, ensuring that the constraints are supported by the framework, and forcing the end user to determine a fitting encoding of their own. In this work, we consider a solution to better support the end user in this task.

\subsection{General Idea}

The solution proposed in this paper rests on the following general idea: When similar problem instances are considered, the corresponding QUBO formulations often only differ in nuances. For example, for pathfinding problems, the input encoding can often be re-used, and only the encoding of constraints has to be adapted between different problem instances. Furthermore, even these constraints are often shared among various problems (e.g., both the TSP and the SOP consider paths that visit all vertices exactly once), and their QUBO formulation can be leveraged multiple times.

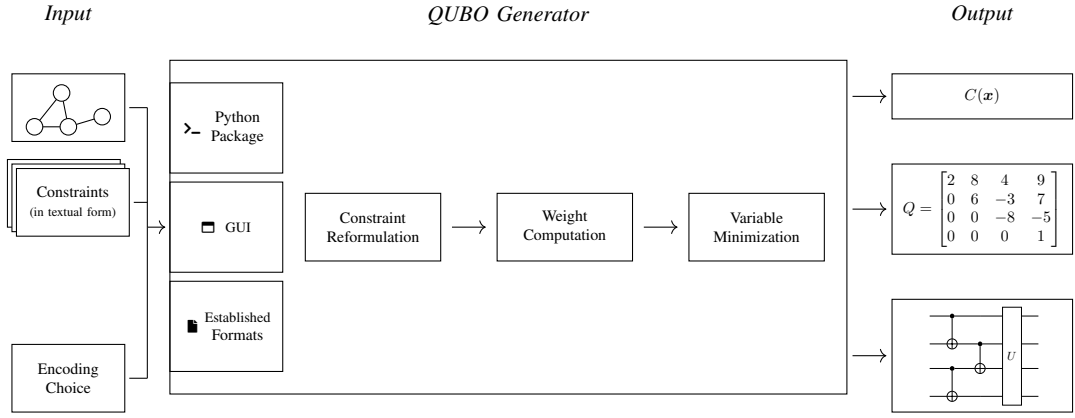
\begin{figure*}
    \centering
    \begin{tikzpicture}[scale=0.6, every node/.style={scale=0.6}]
        \draw (-8.5, -1.3) rectangle (-6, 0.2);
        \begin{scope}[shift={(-7.25,-0.15)}, scale=0.75, every node/.style={scale=0.25}]
            \node[state] (v4) at (-0.2, -0.1) {};
            \node[state] (v1) at (-1, -1.1) {};
            \node[state] (v2) at (0, -1.1) {};
            \node[state] (v3) at (1, -0.8) {};

            \path[-] (v1) edge (v2);
            \path[-] (v2) edge (v3);
            \path[-] (v2) edge (v4);
            \path[-] (v1) edge (v4);
        \end{scope}

        \draw[fill=white] (-8.6, -3.2) rectangle (-6.1, -1.7);
        \draw[fill=white] (-8.5, -3.3) rectangle (-6, -1.8);
        \draw[fill=white] (-8.4, -3.4) rectangle (-5.9, -1.9);
        \node[align=center] at (-7.15, -2.65) {Constraints\\\footnotesize (in textual form)};

        \draw (-8.5, -7.3) rectangle (-6, -5.8);
        \node[align=center] at (-7.25, -6.55) {Encoding\\Choice};

        \coordinate (start1) at (-5.9, -0.55);
        \coordinate (start2) at (-5.8, -2.65);
        \coordinate (start3) at (-5.9, -6.55);

        \coordinate (e1) at (-5.5, -0.55);
        \coordinate (e2) at (-5.5, -2.65);
        \coordinate (e3) at (-5.5, -6.55);
        \coordinate (merge) at (-5.5, -3.2);

        \coordinate (end) at (-5.1, -3.2);

        \draw[-] (start1) -- (e1);
        \draw[-] (start2) -- (e2);
        \draw[-] (start3) -- (e3);
        \draw[-] (e1) -- (merge);
        \draw[-] (e3) -- (merge);
        \draw[->] (merge) -- (end);

        \draw (-5, -6.9) rectangle (10, 0.5);
        \node[align=center, scale=1.4] at (2.5, 1.5) {\emph{QUBO Generator}};

        \draw (-2, -2.45) rectangle (1, -3.95);
        \node[align=center] at (-0.5, -3.2) {Constraint\\Reformulation};
        \draw[->] (1.25, -3.2) -- (2.0, -3.2);

        \draw (2.25, -2.45) rectangle (5.25, -3.95);
        \node[align=center] at (3.75, -3.2) {Weight\\Computation};
        \draw[->] (5.5, -3.2) -- (6.25, -3.2);
        
        \draw (6.5, -2.45) rectangle (9.5, -3.95);
        \node[align=center] at (8, -3.2) {Variable\\Minimization};

        \draw (-5, 0) rectangle (-2.5, -2);
        \node[align=center] at (-4.5, -1) {\footnotesize \faTerminal};
        \node[align=center] at (-3.5, -1) {Python\\Package};
        \draw (-5, -2.2) rectangle (-2.5, -4.2);
        \node[align=center] at (-3.75, -3.2) {\footnotesize \faWindowMaximize[regular] \, \small GUI};

        \draw (-5, -4.4) rectangle (-2.5, -6.4);
        \node[align=center] at (-4.5, -5.4) {\footnotesize \faFile};
        \node[align=center] at (-3.5, -5.4) {\small Established\\Formats};

        \draw (11, -7.3) rectangle (15, -4.8);
        \draw (11, -3.8) rectangle (15, -1.8);
        \node[align=center] (qubo) at (13, -2.8) {$Q = \begin{bmatrix}
            2 & 8 & 4 & 9 \\
            0 & 6 & -3 & 7 \\
            0 & 0 & -8 & -5 \\
            0 & 0 & 0 & 1 \\
        \end{bmatrix} $};
        \draw (11, -0.8) rectangle (15, 0.2);
        \node[align=center] (cx) at (13, -0.3) {$C(\bx)$};

        \node (ocx1) at (10, -0.3) {};
        \node (ocx2) at (11, -0.3) {};
        \node (oqubo1) at (10, -2.8) {};
        \node (oqubo2) at (11, -2.8) {};
        \node (oqc1) at (10, -6.05) {};
        \node (oqc2) at (11, -6.05) {};

        \path[->] (ocx1) edge (ocx2);
        \path[->] (oqubo1) edge (oqubo2);
        \path[->] (oqc1) edge (oqc2);

        \node[scale=0.75] at (13, -6.25) {
            \includegraphics{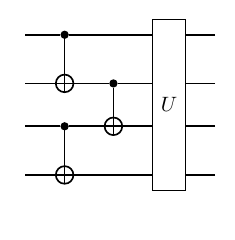}
        };

        \node[align=center, scale=1.4] at (-7.25, 1.5) {\emph{Input}};
        \node[align=center, scale=1.4] at (13, 1.5) {\emph{Output}};
    \end{tikzpicture}
    \caption{The workflow of the proposed framework. The user can use a \emph{GUI} to define their desired constraints or use the \emph{Python package} directly. Alternatively, further input formats such as a \emph{JSON} format or the \emph{TSPLib} format are supported. The output, representing a QUBO formulation of the problem, can then be returned using different formats. These formats can be translated automatically into quantum circuits or classical algorithms to determine the optimal solution to the problem.}
    \label{fig:workflow}
\end{figure*}

Based on these observations, we propose a framework as sketched in Figure~\ref{fig:workflow} to provide a fully automated approach to generate QUBO formulations for almost arbitrary pathfinding problems while avoiding the challenges mentioned earlier. In this representation, problem instances can be solved automatically with the help of existing classical tools and quantum algorithms.

More precisely, this framework will provide the following features:

\begin{enumerate}
    \item \emph{Natural Problem Input Interface}: End users should be shielded as much as possible from the tedious and \mbox{error-prone} steps of formulating QUBO problems. Additionally, the pathfinding problem input formulation should be as easy as possible. Therefore, the proposed framework supports a wide range of problem descriptions, including textual descriptions, existing formats, and GUIs, which allow users to provide the graph and the constraints of their considered problems. Furthermore, all input methods allow the user to easily select and switch among three supported encoding types, giving them the freedom to test and compare the different encodings without any additional work. 
    \item \emph{Automated Constraint Translation}: The framework offers an automatic and mathematically correct translation of the given problem into a QUBO formulation, independent of the number of required constraints defining the problem and the arising complexity from their combination. This is encapsulated in the QUBO Generator module sketched in the middle of Figure~\ref{fig:workflow}.
    \item \emph{Multiple Output Formats}: The resulting QUBO formulation has to be passed to existing classical tools or quantum algorithms to perform the optimization procedure. To this end, the proposed framework supports a large number of output formats. This includes multiple supported output granularities, ranging from a fully prepared quantum circuit to mathematical formulations of cost functions as visualized on the right of Figure~\ref{fig:workflow}.
\end{enumerate}

Following this approach, the proposed framework allows end users without quantum computing knowledge to generate QUBO formulations automatically for the considered problems with low effort, significantly simplifying the usage of quantum computers for this kind of problems.

\section{Proposed Framework}\label{sec:proposed-framework}

In this section, we describe the implementation of the framework sketched above. For more details and for actually trying out the framework, we refer to the open-source implementation on \href{https://github.com/cda-tum/mqt-qubomaker}{https://github.com/cda-tum/mqt-qubomaker}, as well as the supporting GUI accessible through a web interface \href{https://cda-tum.github.io/mqt-qubomaker/}{https://cda-tum.github.io/mqt-qubomaker/}. To this end, we focus on the core concepts of the \emph{classical input interface}, the \emph{automated translation} process, and the generated \emph{output formats}.

\subsection{Natural Problem Input Interface}
\label{sec:interface}

This section defines the inputs involved during the workflow of the proposed framework, shown in Figure~\ref{fig:workflow}. As reviewed in Section~\ref{sec:background}, the key components for the construction of a QUBO formulation for pathfinding problems are:
\begin{itemize}
    \item \emph{Problem Graph}: The graph considered for the individual problem instance.
    \item \emph{Constraints}: The constraints required to correctly identify the specific problem, in natural language, unrelated to their actual QUBO encodings.
    \item \emph{Encoding Choice}: The encoding to be used for the representation of the found paths, as a choice among three supported encodings.
\end{itemize}

As mentioned above, the proposed framework aims to minimize the work required for the user to define and reformulate their specific problems, shielding them from the knowledge required outside their expertise. To this end, it supports multiple interfaces for user interaction (examples are presented later in Section~\ref{sec:casestudy}):

\begin{enumerate}

\item \emph{Python Library}: As a Python library, end users can directly access the framework's core functionality through the specialized classes for QUBO generation. The library provides a module that contains classes for all supported constraints that can be instantiated and added to the QUBO-generation object.

\item \emph{GUI}: The framework also provides a graphical user interface that allows the end user to define constraints and problem settings in a no-code fashion.

\item \emph{Established Formats}: Finally, the proposed framework also supports established input formats. This allows end users to provide problem definitions in formats they are already familiar with, drastically reducing the entry barriers. Using the \emph{JSON~format}, the user can provide a JSON file to define the problem settings and all required constraints through the corresponding method for parsing problem definitions. Alternatively, this framework supports a subset of \emph{TSPLib}~\cite{reinelt1991} specification files. This format, regularly used to define several pathfinding problems for benchmarking purposes, can be translated to a QUBO problem formulation directly, allowing users to provide their \emph{TSPLib} inputs directly without requiring any form of modification.

\end{enumerate}

This broad range of input formats allows end users to select the format with which they feel the most comfortable, significantly reducing the hurdle for the creation of QUBO formulations.

\subsection{Automated Constraint Translation}
\label{sec:constraints}

Once an input defining the problem instance has been passed to the framework, it can generate a QUBO formulation for the problem automatically by translating and combining all individual constraints. This process requires a valid mathematical formulation for each combination of constraints and encodings. 

The following describes the corresponding translation process. To this end, the particular constraints are explored for a given graph $G = (V, E)$. Further, $N$ represents the maximum path length, $\Pi$ the set of all paths to be found for the solution of the problem, and $\pipath^{(i)}$ the $i$-th of these paths. 

Having that as a basis, a key concept for the construction of QUBO formulations is the definition of an encoding. While encodings can, in general, be defined specifically for particular problems, using general encodings has the advantage of a broader range of compatibility with individual problems. The specific encoding defines not only the number of binary variables required for the computation of the solution but also the complexity of the corresponding cost functions and the convergence behavior, depending on the employed algorithm. Therefore, supporting multiple different encoding types has the advantage of giving the approach more adaptability. Problems that cannot be solved efficiently using one encoding may, therefore, find solutions more easily in a different encoding. Through the proposed framework, the end user can easily test and evaluate any of the supported encodings without the tedious reformulation process usually required for this task.

In the current version, the proposed framework supports three path-encoding types that each have individual advantages and disadvantages:

\begin{itemize}
\item \emph{One-Hot Encoding} This encoding provides a set of ${N \cdot |V| \cdot |\Pi|}$ binary variables $x_{v, j, \pipath^{(i)}}$. An individual variable with value $1$ indicates that the corresponding vertex~$v$ is located at position $j$ in path $\pipath^{(i)}$. Assignments such that $x_{v, j, \pipath^{(i)}} = x_{v', j, \pipath^{(i)}} = 1$, with $v \neq v'$  are invalid.
\item \emph{Domain Wall Encoding} This encoding provides a set of ${N \cdot |V| \cdot |\Pi|}$ binary variables $x_{v, j, \pipath^{(i)}}$. For a given vertex~$v$ to be located at position $j$ in path $\pipath^{(i)}$, the encoding requires $x_{u, j, \pipath^{(i)}} = 1$ for all $u \leq v$. Compared to the one-hot encoding, which uses the same number of variables, this encoding has the advantage of allowing the transition between two valid states with just one bitflip, while the one-hot encoding requires two bitflips~\cite{chancellor2019}.
\item \emph{Binary Encoding} This encoding uses ${\text{log}N \cdot |V| \cdot |\Pi|}$ binary variables. For a given vertex~$v$ to be located at position $j$ in path $\pipath^{(i)}$, the bitstring given by $x_{u, j, \pipath^{(i)}}$, where $u \in V$, has to be the binary representation of~$v$. The number of binary variables required for this encoding is asymptotically lower than that of the previous encodings, and there are no invalid assignments, at the cost of requiring more involved cost functions due to the complex nature of the encoding.
\end{itemize}

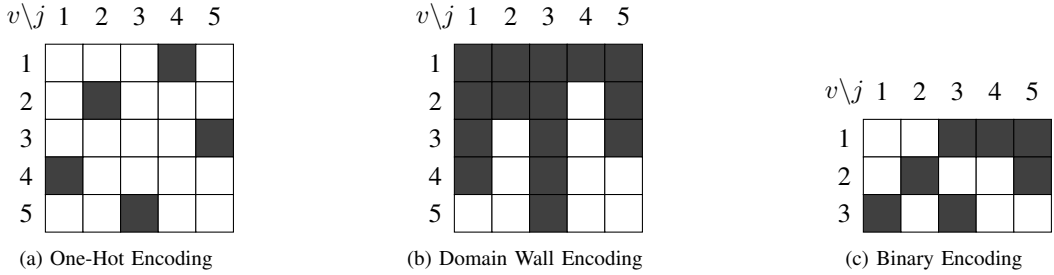
\begin{figure*}
    \centering
    \begin{subfigure}{0.6\columnwidth}
        \centering
        \begin{tikzpicture}[scale=0.5]
            \def\matrixContent{{0,0,0,1,0},
                {0,1,0,0,0},
                {0,0,0,0,1},
                {1,0,0,0,0},
                {0,0,1,0,0},
            }
            \foreach \row [count=\y] in \matrixContent {
                \foreach \value [count=\x] in \row {
                    \ifnum \value=0
                        \node[draw, rectangle, minimum size=0.5cm, fill=white] at (\x + 1,-\y) {};
                    \else
                        \node[draw, rectangle, minimum size=0.5cm, fill=darkgray] at (\x + 1,-\y) {};
                    \fi
                }
            }
            \foreach \x in {1,...,5} {
                \node at (\x + 1,0.25) {\x};
                \node at (1,-\x) {\x};
            }
            \node at (1, 0.25) {$v \backslash j$};
        \end{tikzpicture}
        \caption{One-Hot Encoding}
        \label{fig:encoding-one-hot}
    \end{subfigure}
    \begin{subfigure}{0.6\columnwidth}
        \centering
        \begin{tikzpicture}[scale=0.5]
            \def\matrixContent{{1,1,1,1,1},
                {1,1,1,0,1},
                {1,0,1,0,1},
                {1,0,1,0,0},
                {0,0,1,0,0},
            }
            \foreach \row [count=\y] in \matrixContent {
                \foreach \value [count=\x] in \row {
                    \ifnum \value=0
                        \node[draw, rectangle, minimum size=0.5cm, fill=white] at (\x + 1,-\y) {};
                    \else
                        \node[draw, rectangle, minimum size=0.5cm, fill=darkgray] at (\x + 1,-\y) {};
                    \fi
                }
            }
            \foreach \x in {1,...,5} {
                \node at (\x + 1,0.25) {\x};
                \node at (1,-\x) {\x};
            }
            \node at (1, 0.25) {$v \backslash j$};
        \end{tikzpicture}
        \caption{Domain Wall Encoding}
        \label{fig:encoding-unary}
    \end{subfigure}
    \begin{subfigure}{0.6\columnwidth}
        \centering
        \begin{tikzpicture}[scale=0.5]
            \def\matrixContent{
                {0,0,1,1,1},
                {0,1,0,0,1},
                {1,0,1,0,0},
            }
            \foreach \row [count=\y] in \matrixContent {
                \foreach \value [count=\x] in \row {
                    \ifnum \value=0
                        \node[draw, rectangle, minimum size=0.5cm, fill=white] at (\x + 1,-\y) {};
                    \else
                        \node[draw, rectangle, minimum size=0.5cm, fill=darkgray] at (\x + 1,-\y) {};
                    \fi
                }
            }
            \foreach \x in {1,...,5} {
                \node at (\x + 1,0.25) {\x};
            }
            \foreach \x in {1,...,3} {
                \node at (1,-\x) {\x};
            }
            \node at (1, 0.25) {$v \backslash j$};
        \end{tikzpicture}
        \caption{Binary Encoding}
        \label{fig:encoding-binary}
    \end{subfigure}
    \caption{Three different encodings representing the same Hamiltonian cycle on the graph shown in Figure~\ref{fig:graph}.}
    \label{fig:encodings}
\end{figure*}

\begin{example}
Figure~\ref{fig:encodings} presents the example path~$(v_4, v_2, v_5, v_1, v_3)$ in the three different encodings as tables, where the rows represent the index $v$ and the columns represent the index $j$ of $x_{v, j, \pipathi}$. Here, it is assumed that $|\Pi| = 1$ and, consequently, the index $\pipath^{(i)}$ is omitted for the sake of presentation.
    
\end{example}

Once the end user selects the desired encoding through the user interface, the required constraints can be translated automatically into QUBO form before being combined into a single problem-specific cost function. For this matter, the framework provides support for a total of twelve constraints with pre-defined QUBO formulations for each encoding:

\begin{enumerate}
        \item \label{constr:valid} $\pipathi$ represents a valid path: Enforces that edges exist between any pair of consecutive vertices $v_{i_j}, v_{i_{j+1}}$ and further enforces that the assignment follows the rules of the selected encoding.
        \item \label{constr:pos} The $j$-th element of $\pipathi$ is one of $\{v_1, v_2, ..., v_k\}$.
        \item All \emph{vertices} in a given set of vertices appear \emph{at~least~once} in~$\pipathi$.
        \item \label{constr:exactly} All \emph{vertices} in a given set of vertices appear \emph{exactly once} in $\pipathi$.
        \item All \emph{vertices} in a given set of vertices appear \emph{at~most~once} in~$\pipathi$.
        \item All \emph{edges} in a given set of edges appear \emph{at~least~once} in~$\pipathi$.
        \item All \emph{edges} in a given set of edges appear \emph{exactly once} in~$\pipathi$.
        \item All \emph{edges} in a given set of edges appear \emph{at~most~once} in~$\pipathi$.
        \item \label{constr:share-vertices} Given a set of paths, no two individual paths of the set \emph{share vertices}.
        \item Given a set of paths, no two individual paths of the set \emph{share edges}.
        \item \label{constr:prec} Given two vertices $u, v$, enforce a \emph{precedent constraint}: vertex $v$ may not appear in any path before visiting vertex $u$.
        \item \label{constr:opt} The total weight of all paths in a given set of paths is \emph{is minimal} / \emph{maximal}.
\end{enumerate}

In addition to that, the proposed framework also supports constraints that cannot be represented by quadratic cost functions (addressing another challenge discussed in Section~\ref{sec:challenges}): depending on the complexity of the constraint and the selected encoding, some cost functions require higher-order products to be formulated. In these cases, the framework automatically extends the cost function by adding auxiliary variables. These auxiliary variables are created with the potential of reuse in mind. For each auxiliary variable, the framework stores the partial product it represents, and at later stages, higher-order products are first searched for partial products that have been encountered before and replaced by these variables. This way, the number of required auxiliary variables stays minimal.

\begin{example}
    Consider the binary cost function ${C(\bx) = x_1x_2 - 2x_3 + x_1x_2x_3}$. The terms $x_1x_2$ and $-2x_3$ are of quadratic and linear order, respectively, and, thus, are valid in the context of QUBO formulations. The term~$x_1x_2x_3$, however, involves three variables and has to be reformulated. This can be achieved by introducing the auxiliary variable $y$. By adding the correct penalty terms, the equality $y = x_1x_2$ can be enforced, replacing $x_1x_2x_3$ with $x_1x_2 - 2x_1y - 2x_2y + 3x_3y + 3y$. By using this approach, the order of the term was decreased by one, and it is now valid for a QUBO formulation. To reformulate terms of order 4 or higher, this strategy can be employed repeatedly, increasing the number of required variables by 1 for each step.
\end{example}

As the task of predicting the number of required auxiliary variables for a particular problem instance on a given graph and with a pre-defined encoding is not trivial, the framework further supplies a utility to assist in the selection of encodings. For a given problem definition, it compares all supported encodings and evaluates which encoding requires the lowest number of binary variables for its QUBO formulation. This allows the user to easily select the most efficient encoding without requiring an understanding of any of the mathematical foundations.

\subsection{Output Formats}

After the required input has been provided to the framework and the automated QUBO formulation process is complete, one of several levels of output granularities can be selected for the representation of the resulting formula:

\begin{itemize}
    \item As a single \emph{Binary Cost Function}: A purely mathematical representation of the QUBO problem as a function $C(\bx)$ over the binary variables $\bx$.
    \item As a \emph{QUBO Matrix}: A triangular matrix $Q$ that uniquely represents the QUBO problem as an optimization problem of the form $\argmin_{\bx} \bx^TQ\bx$.
    \item As a \emph{Hamiltonian operator}: A quantum operator based on Qiskit that, when measured, yields the value of the corresponding QUBO cost function as its expectation value.
    \item As a \emph{QAOA quantum circuit}: A Qiskit implementation of the QAOA algorithm, defined for the specific problem instance. Further QAOA parameters, such as repetition count, can be specified by the user.
\end{itemize}

This broad range of different output formats ensures a large variety of application areas. Not only can this framework be used to generate quantum computing solutions to pathfinding problems that are ready to be used on real quantum devices, but it also supports the generation of the QUBO formulation in its mathematical form so that it can be used with classical QUBO solvers, or as a Hamiltonian operator, so that it may be coupled with different quantum algorithms. In all cases, the end user is provided with a QUBO formulation that has been generated automatically and did not require any QUBO or quantum computing background.

\section{Case Studies}\label{sec:casestudy}

\begin{figure*}
    \centering
    \includegraphics[width=\textwidth]{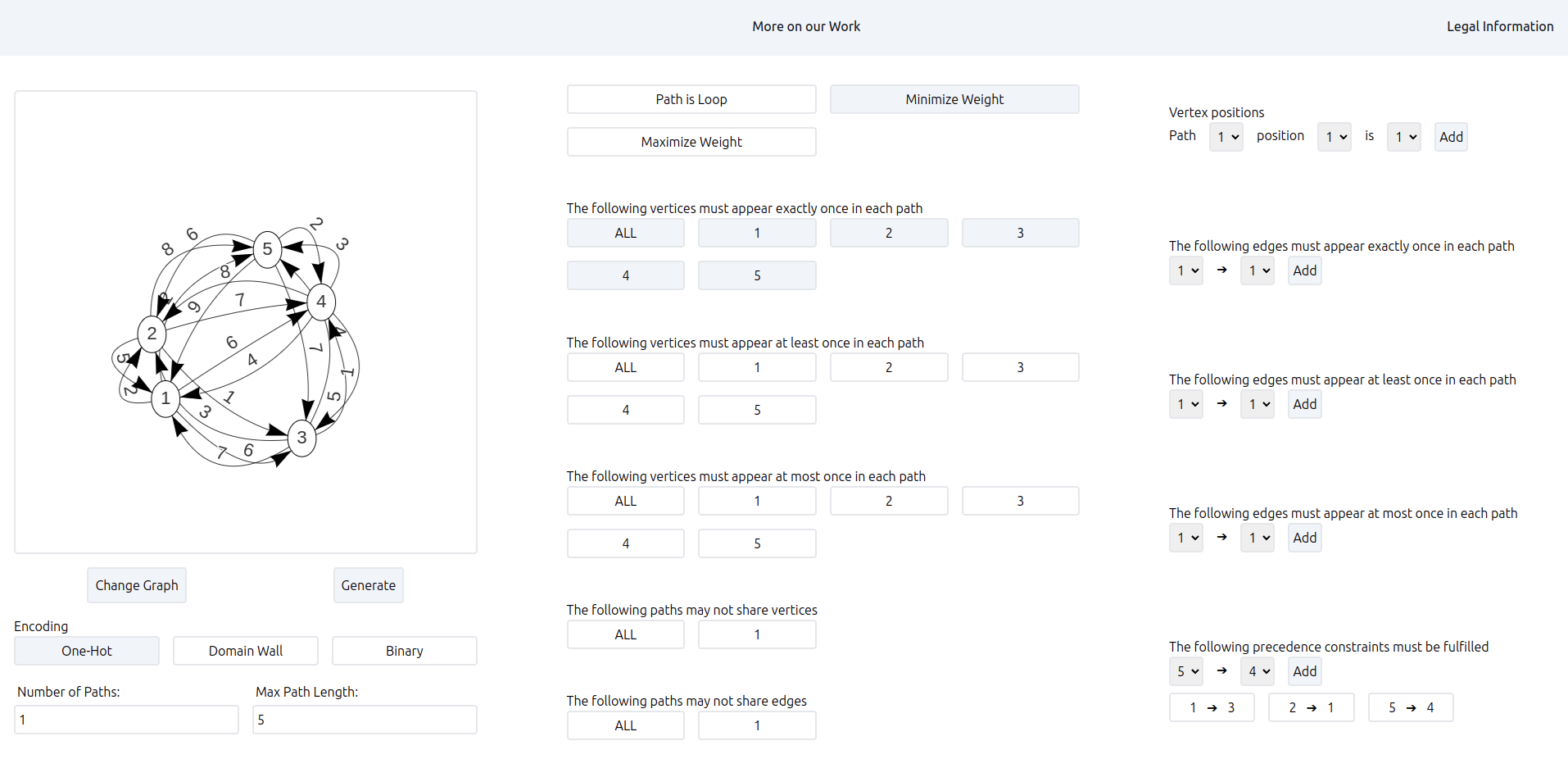}
    \caption{User interface of the pathfinding tool with inputs specifying an SOP instance. The left column visualizes the graph considered for the current problem instance, as well as settings such as encoding choice. The middle and right columns allow for the selection of a variety of constraints that will be added to the final QUBO formulation. Pressing the \emph{"Generate"} button will transform the selected settings and constraints into an input for the proposed framework.}
    \label{fig:screenshot-gui}
    \vspace*{-0.35cm}
\end{figure*}

In order to demonstrate the accessibility and utility of the proposed framework in solving various pathfinding problems, to confirm that the challenges discussed in Section~\ref{sec:challenges} are addressed by it, and to evaluate the three different encoding schemes, several case studies on different problems have been conducted. In this section, the correspondingly obtained results are summarized. To this end, we first outline two of the considered case studies. Afterward, we discuss the results of these case studies, compare their performance among the different encoding schemes, and highlight key differences in the workflow with existing tools with similar goals. 

The correspondingly used open-source implementations and GUI are available through \href{https://github.com/cda-tum/mqt-qubomaker}{https://github.com/cda-tum/mqt-qubomaker} and \href{https://cda-tum.github.io/mqt-qubomaker/}{https://cda-tum.github.io/mqt-qubomaker/}, respectively.

\subsection{Case Study: Sequential Ordering Problem}

In a first case study, we considered the \emph{Sequential Ordering Problem (SOP)} which is defined by a graph $G = (V, E)$ and a set of \emph{ordering constraints}~${O = \{(v_{a_1} \prec v_{b_1}), ..., (v_{a_k} \prec v_{b_k})\}}$. It has the goal of finding the lowest-weight path $\pipath$ that visits all vertices in $V$ exactly once while abiding by the ordering constraints: for each $(v_{a_i} \prec v_{b_i}) \in O$, vertex $v_{b_i}$ may not be visited before visiting vertex $v_{a_i}$. 

Given an SOP instance based on the graph shown in Figure~\ref{fig:graph} and the ordering constraints~${O = \{(v_1 \prec v_3), (v_2 \prec v_1), (v_5 \prec v_4)\}}$, the problem is translated into its QUBO formulation using the proposed framework by breaking it down into the following constraints (denoted in brackets) from Section~\ref{sec:constraints}, i.e.,
\begin{itemize}
    \item (\ref{constr:exactly}) all vertices $v \in V$ appear exactly once in $\pipath$,
    \item (\ref{constr:prec}) $v_1$ appears before $v_3$ in $\pipath$,
    \item (\ref{constr:prec}) $v_2$ appears before $v_1$ in $\pipath$,
    \item (\ref{constr:prec}) $v_5$ appears before $v_4$ in $\pipath$, and
    \item (\ref{constr:opt}) $w(\pipath)$ is minimal.
\end{itemize}

These constraints are passed to the proposed framework directly through its Python package or GUI. Figure~\ref{fig:screenshot-gui} shows an example configuration of the user interface for this problem. On the left, the problem graph is displayed, passed to the interface through its adjacency matrix, and the settings for the QUBO formulation are specified. In this case, the \mbox{\emph{One-Hot}} encoding is selected. Furthermore, the \emph{"Minimize Weight"} constraint has been chosen, as well as the requirement for each vertex to \emph{appear exactly once} in the path. Finally, on the bottom right, the three \emph{ordering constraints} for this problem instance have been provided. Pushing the \emph{"Generate"} button now creates a JSON-encoded specification of the problem instance that can be passed to the framework to directly generate a QUBO formulation for this specific problem instance. Figure~\ref{fig:eval-qubo-sop} sketches the correspondingly obtained result as a Hamiltonian. Note that this QUBO formulation was constructed without any need for decisions on binary variables or any mathematical expressions based on them, with the input based solely on the user's expertise on the specific pathfinding problem.

As an alternative approach, SOP problem instances can also be provided through the framework's \emph{TSPLib} interface. This popular format for pathfinding problem specification allows the definition of various problem instances, including SOP. By passing a \emph{TSPLib} specification of this problem to the proposed framework, its QUBO formulation can be constructed directly without any further user input.

Now, the QUBO formulation can be used as a basis for optimization algorithms to compute the optimal solution. By generating the output as a Hamiltonian operator, as shown in Figure~\ref{fig:eval-qubo}, it can be used directly by different quantum algorithms to perform the optimization.

\subsection{Case Study: Two Disjoint Paths Problem}

In a second case study, we considered the \emph{Two Disjoint Paths \mbox{Problem (2-DPP)}} which is given by a graph ${G = (V, E)}$, two starting vertices $v_{s_1}$, $v_{s_2}$ and two target vertices $v_{t_1}$, $v_{t_2}$. The optimal solution to this problem is two paths, $\pipathN{1}$ and $\pipathN{2}$ with minimal total weight, starting and ending at $v_{s_1}$, $v_{t_1}$ and $v_{s_2}$, $v_{t_2}$, respectively such that the paths do not intersect on any vertex (i.e., $V(\pipathN{1}) \cap V(\pipathN{2}) = \emptyset$ with $V(\pipath)$ representing the vertices visited by path $\pipath$).

This example considers, once again, the graph shown in Figure~\ref{fig:graph}, with $(v_{s_1}, v_{t_1}) = (v_1, v_4)$ and $(v_{s_2}, v_{t_2}) = (v_2, v_5)$. Now, the QUBO formulation of this problem can be generated by the proposed framework by dividing it into the constraints, i.e.,
\begin{itemize}
    \item (\ref{constr:pos}) $\pipathN{1}$ starts at $v_1$,
    \item (\ref{constr:pos}) $\pipathN{2}$ starts at $v_2$,
    \item (\ref{constr:pos}) $\pipathN{1}$ ends at $v_4$,
    \item (\ref{constr:pos}) $\pipathN{2}$ ends at $v_5$,
    \item (\ref{constr:share-vertices}) $\pipathN{1}$ and $\pipathN{2}$ do not share any vertices, and
    \item (\ref{constr:opt}) $w(\pipathN{1}) + w(\pipathN{2})$ is minimal.
\end{itemize}

Once again, the constraints can be passed directly to the framework to generate the QUBO formulation. In this case, the GUI can be configured by setting the desired \emph{"Number of Paths"} to 2. Furthermore, \emph{"Minimize Weight"} is selected once more, and paths 1 and 2 are selected as \emph{not sharing vertices} on the bottom of the middle column. Finally, the starting and target vertices are included in the \emph{"Vertex Positions"} constraints. An excerpt of the resulting Hamiltonian is shown in Figure~\ref{fig:eval-qubo-2pp}. While this problem instance requires constraints that are very different from the ones proposed for the SOP, it still does not involve any additional work to adapt the framework to the 2-DPP, as all required constraints are already supported by the framework.

Similarly to before, this formulation can then be used as the input for the quantum algorithms mentioned

\begin{figure}
    \centering
    \begin{subfigure}{\columnwidth}
        \centering
        \[\begin{bmatrix}
            0 & 92 & 92 & 92 & \cdots & 0 \\
            0 & -46 & 92 & 92 & \cdots & 0 \\
            0 & 0 & 0 & 92 & \cdots & 0 \\
            0 & 0 & 0 & 0 & \cdots & 0 \\
            \vdots & \vdots & \ddots & \vdots \\
            0 & 0 & 0 & 0 & \cdots & 138
        \end{bmatrix}\]
        \caption{SOP}
        \label{fig:eval-qubo-sop}
    \end{subfigure}
    \begin{subfigure}{\columnwidth}
        \centering
        \[\begin{bmatrix}
            0 & 56 & 56 & 224 & \cdots & 0 \\
            0 & 28 & 56 & 201 & \cdots & 0 \\
            0 & 0 & 28 & 147 & \cdots & 0 \\
            0 & 0 & 0 & 4 & \cdots & 0 \\
            \vdots & \vdots & \ddots & \vdots \\
            0 & 0 & 0 & 0 & \cdots & 168
        \end{bmatrix}\]
        \caption{2-DPP}
        \label{fig:eval-qubo-2pp}
    \end{subfigure}
    \caption{Excerpts of matrix representations of the Hamiltonian operators generated by the QUBO generator for the problem instances in this evaluation.}
    \label{fig:eval-qubo}
    \vspace*{-0.25cm}
\end{figure}

\subsection{Discussion}

As shown by the two case studies, the burden on the end user for the construction of QUBO formulations has been drastically reduced in comparison to the solution using existing tools as reviewed in Section~\ref{sec:related-work} and illustrated in Example~\ref{ex:4}:
\begin{enumerate}
    \item No expertise in QUBO problems is required. Figure~\ref{fig:screenshot-gui} clearly confirms that the problem can be described in the actual problem domain, while the end user is completely shielded from tasks involving QUBO or quantum computing concepts.
    \item No mathematical expressions have to be created manually, avoiding a common source of errors and giving the user confidence in the generated QUBO formulation, even in complex cases.
    \item The end user no longer needs to be concerned about the exact form of individual cost functions. Reformulations required for higher-order terms are performed automatically. This way, the user can select any supported constraint without the need to make sure it can be expressed using quadratic terms.
    \item The end user is not required to manually merge all constraints into a single cost function, as this process is part of the framework's automated workflow.
\end{enumerate}

This clearly confirms the accessibility and utility of the proposed framework and, indeed, addresses all the challenges discussed in Section~\ref{sec:challenges}.

While aspects (1) and (4) were already achieved through the help of existing tools, as illustrated in Example~\ref{ex:4}, the framework proposed in this work completely frees the end user from any form of mathematical reformulation of constraints. On the one hand, this sacrifices the open nature of the mathematical constraint programming approach, which can be employed to solve many different optimization problems in various fields. On the other hand, it can guarantee correct solutions in a simple and automatic fashion for almost arbitrary pathfinding problems. All tasks required from the user are based solely on this domain, greatly simplifying the process of creating and modifying QUBO formulations.

Moreover, the proposed framework provides value through its support of different encodings. This allows the user to easily evaluate individual QUBO formulations, switching between encodings at the click of a button, even as the resulting output changes disproportionately to the required effort. In comparison, even with the help of existing tools, a change of encoding requires users to redefine all constraints and their mathematic expressions, effectively requiring them to start the task from scratch for each new encoding.

In particular, the performance of individual encodings can be evalauted based on multiple factors:
\begin{itemize}
    \item \emph{Variable Count}: Due to the nature of each encoding scheme, the number of variables required for each of them differs depending on individual scenarios. While the \emph{Binary} encoding, for instance, requires the lowest number of variables to encode a given path, its complexity often requires higher-order terms to formulate cost functions and, thus, may lead to a large number of auxiliary variables. On the other hand, the \emph{Domain Wall} and \mbox{\emph{One-Hot}} encoding schemes often require just very simple cost functions, leading to a lower number of required auxiliary variables.
    \item \emph{Convergence Behaviour}: As the encoding scheme directly correlates to the interpretation of a given variable assignment, optimization algorithms can pass through the optimization landscapes at different speeds depending on the chosen encoding. For instance, the \emph{Binary} encoding does not lead to any invalid assignments. Any assignment can always be interpreted as a valid path. In the \mbox{\emph{One-Hot}} encoding, on the other hand, the number of correct assignments is exponentially small compared to possible invalid assignments. This can impact both the convergence time as well as the likelihood of reaching the globally optimal solution.
\end{itemize}

\begin{table}[]
    \caption{The number of required variables for the case study problems.}
    \label{tab:variable-counts}
    \centering
    \renewcommand{\arraystretch}{1.5}
    \begin{tabular}{l c c}
        \hline
        Encoding & SOP & 2-DPP \\ \hline
        \emph{One-Hot} & 58 & 112 \\ \hline
        \emph{Domain Wall} & 183 & 30 \\ \hline
        \emph{Binary} & 2375 & 160 \\
    \end{tabular}
    \vspace*{-0.25cm}
\end{table}

Table~\ref{tab:variable-counts} illustrates the variable requirements of both problems considered in the case studies above for all supported encoding schemes. This clearly shows the volatility of the encodings and the drastic improvements in variable count that can be achieved by selecting a different encoding. In the case of \mbox{2-DPP}, the \emph{Domain Wall} encoding did not require any auxiliary variables, as all higher-order terms were canceled out due to the nature of the encoding. Contrarily, in the case of SOP, the \emph{One-Hot} encoding managed to require the lowest amount of variables, while the complexity of the \emph{Binary} encoding resulted in more than 2000 auxiliary variables.

\revised{Thus far, evaluations like these have been a cumbersome and time-consuming task. While comparisons of different encoding schemes based on their performance exist for specific problem instances~\cite{chen2021, codognet2022, berwald2022, sawaya2020, chancellor2019}, the task of manually rewriting QUBO formulations for different encodings has proven to be too inefficient for thorough evaluation in most cases. In fact, once an end user struggled to get to one encoding and its corresponding QUBO formulation, they often did not bother to go through this hassle again for other encodings. Contrarily, using the proposed framework, different encodings can be evaluated at the click of a button.}

\section{Conclusion}\label{sec:conclusion}

This work proposed an automatic approach to the construction of QUBO formulations from almost arbitrarily defined pathfinding problems. By generating binary cost functions directly from problem constraints on a conceptual level, the proposed framework is able to shield end users from any required experience in the area of QUBO problems or quantum computing.

The proposed framework allows end users to select from a large number of different constraints and easily switch among three supported encodings, allowing them to investigate the performance of each encoding individually without requiring the QUBO formulation to be rewritten. Circumventing problems encountered both in the manual construction of QUBO formulations as well as the application of existing tools, this approach protects users from otherwise tedious and \mbox{error-prone} reformulation tasks, guaranteeing sound and qualitative results that can be passed directly to other frameworks and algorithms, including classical QUBO optimization tools as well as quantum algorithms such as QAOA or VQE.

\revised{The selection of encodings and constraints is supported through a Python package and a graphical user interface that both allow end users to define their problem instances in a semantic way, independent of the underlying mathematical nature of their related cost functions. The open-source framework is available through GitHub oh \href{https://github.com/cda-tum/mqt-qubomaker}{https://github.com/cda-tum/mqt-qubomaker} and has a graphical user interface at \href{https://cda-tum.github.io/mqt-qubomaker/}{https://cda-tum.github.io/mqt-qubomaker/}.}

In future work, the proposed framework can be expanded in multiple directions. Currently, three general encoding schemes are supported. However, the literature lists many more encoding schemes with different advantages or that are specialized for specific problem instances. These encoding schemes can be included in the proposed framework, increasing its adaptability, as a larger number of supported encodings allows users to choose the optimal encoding with more freedom. As the selected encoding schemes can be switched without any overhead for the end user, such improvements would be a powerful addition for evaluation purposes.

Additionally, the application area can be extended by supporting further classes of problems. Currently, the framework focuses on pure pathfinding problems only. However, it is in no way restricted to just that area, and room for automation exists in many other classes of optimization problems, such as network flow problems, scheduling problems, etc. The proposed framework is implemented in a modular fashion, making the addition of further problem classes an approachable task.

Finally, the interaction methods of the framework may further be adapted to target a larger number of users. While support for established formats, such as \emph{TSPLib}, is already implemented, additional existing formats may also be included. Alternatively, natural language processing models may also be employed to interpret problem descriptions in purely natural language, allowing end users to create QUBO formulations without even deconstructing their problem instances into the corresponding constraints. This effort would increase the accessibility of the proposed framework, further decreasing the expertise required to construct QUBO formulations.

Overall, this work proposed a powerful starting step to the effort of making the use of quantum computers and their optimization algorithms more accessible to users without expertise in quantum computing or related fields while also providing tools to aid quantum computing experts in the construction of correct and efficient problem formulations. With the advancements in quantum algorithms for optimization problems made in recent years making them more and more attractive to a wide range of users, it addresses an important issue in the adoption of quantum computing.

\section*{Acknowledgments}
This work received funding from the European Research Council (ERC) under the European Union’s
Horizon 2020 research and innovation program (grant agreement No. 101001318), was part of the
Munich Quantum Valley, which is supported by the Bavarian state government with funds from the
Hightech Agenda Bayern Plus, and has been supported by the BMWK on the basis of a decision by
the German Bundestag through project QuaST, as well as by the BMK, BMDW, the State of Upper
Austria in the frame of the COMET program, and the QuantumReady project within Quantum
Austria (managed by the FFG).

\printbibliography

\end{document}